\documentclass{PoS}

\title{Radio Properties of Brightest Cluster Members}

\ShortTitle{Radio Properties of Brightest Cluster Members}

\author{\speaker{Heinz Andernach}$^{~a, b}$ and Miriam E. Ramos-Ceja~$^{b}$\\
       \llap{$^a$}Argelander-Institut f\"ur Astronomie (AIfA), Universit\"at Bonn, Auf dem H\"ugel 71\\
            D-53121 Bonn, Germany; on leave of absence from: \\
       \llap{$^b$}Departamento de Astronom\'{\i}a, Universidad de Guanajuato, AP 144\\
            Guanajuato, CP 36000, Gto, Mexico\\
       E-mail: \email{heinz@astro.uni-bonn.de},~\email{miriam@astro.ugto.mx}}

\abstract{
We searched the literature for radio continuum images and flux densities of 
the brightest cluster members (BCMs) in 1169 Abell clusters.
The clusters were selected on the basis of their morphological type 
(Bautz-Morgan or Rood-Sastry) or on textual notes in the Abell catalog
indicating the presence of dominant galaxies. We inspected the images of existing 
radio surveys (NVSS, SUMSS, FIRST, WENSS, etc.) and used the CATS and VizieR
catalog browsers, as well as additional literature, to collect radio fluxes and 
radio morphologies for 1423 BCMs in these clusters. We found 578 (41\,\%) of these
BCMs with detected radio emission, of which 223 are detected at only
a single frequency, usually at 1.4~GHz (NVSS or FIRST) or 843~MHz (SUMSS).
Using the survey images and additional published high-resolution images, we 
estimated the best available position angle for the innermost radio structure 
and for the largest angular size of each source. 
Digitized Sky Survey images were used to obtain the orientation of the optical
major axis of the outer envelope of the BCMs, and the acute difference 
angle between major optical and radio axes was derived for 102 objects. 
Its distribution shows a similar bimodality as reported previously for a 
larger BCM sample, although KS tests on both do not distinguish them from
uniform ones. The shape of the distribution is independent of optical 
ellipticity, optical morphological type and largest linear radio size of the 
objects. In order to find further clusters with dominant central galaxies, the 
above-mentioned criteria need to be relaxed.
From the present study there is at most marginal evidence for a relation
between powerful high-redshift radio galaxies and dominant galaxies of
low-$z$ clusters when comparing their radio-optical alignment angle.
We plan to derive radio spectra, radio luminosities and search for
possible relations between these parameters and the radio-optical
alignment, as well as to quantify the radio morphology and search for
relations with the peculiar velocity of the BCMs in their host cluster.
}

\FullConference{Panoramic Radio Astronomy: Wide-field 1-2 GHz research on galaxy evolution\\
                June 2-5 2009\\
                Groningen, the Netherlands}

\begin{document}

\section{Introduction}

Powerful radio galaxies at high redshifts ($z\gtrsim 0.5$)
tend to show a strong alignment of their radio and optical major axes, and
observations support two models for the origin of this effect \cite{MB08}:
jet$-$induced star formation and scattering by dust of the light of a
central AGN. If these objects were the precursors of dominant cD$-$like
galaxies in low-$z$ clusters, as suggested by \cite{DJOR87,MB08}, 
these latter galaxies should also show signs of a radio$-$optical alignment 
\cite{WEST94}. 
Even though the general population of nearby elliptical radio
galaxies favors a radio axis closer to the optical minor axis (cf.\ \cite{AND95}
for references), a small population of aligned radio galaxies was found among
cluster-dominant galaxies \cite{AND95}. This suggests that environment
may influence the relative orientation of optical and radio axes.
Here we describe an attempt to increase previous samples based on a
systematic search for available radio data of a large sample of
optically selected Abell clusters.

\section{Methodology}

The sample of brightest cluster members (BCMs) was taken from
\cite{COZIOL09} with very minor changes, and consists of 1423 galaxies
in 1221 redshift components (i.e.\ clusters along the line of sight)
towards 1169 different Abell clusters. The latter include the supplementary
S-clusters and were selected by their morphological type (Bautz \& Morgan
(BM) I or I$-$II, Rood \& Sastry (RS) cD), and by comments in the Abell
catalog \cite{ACO89} that suggest the presence of a cD galaxy or a ``corona''
around the BCM.

\begin{figure}[t!]
\vspace*{-3mm}
\begin{center}
\includegraphics[width=0.9\textwidth,,viewport=36 120 482 395,clip]{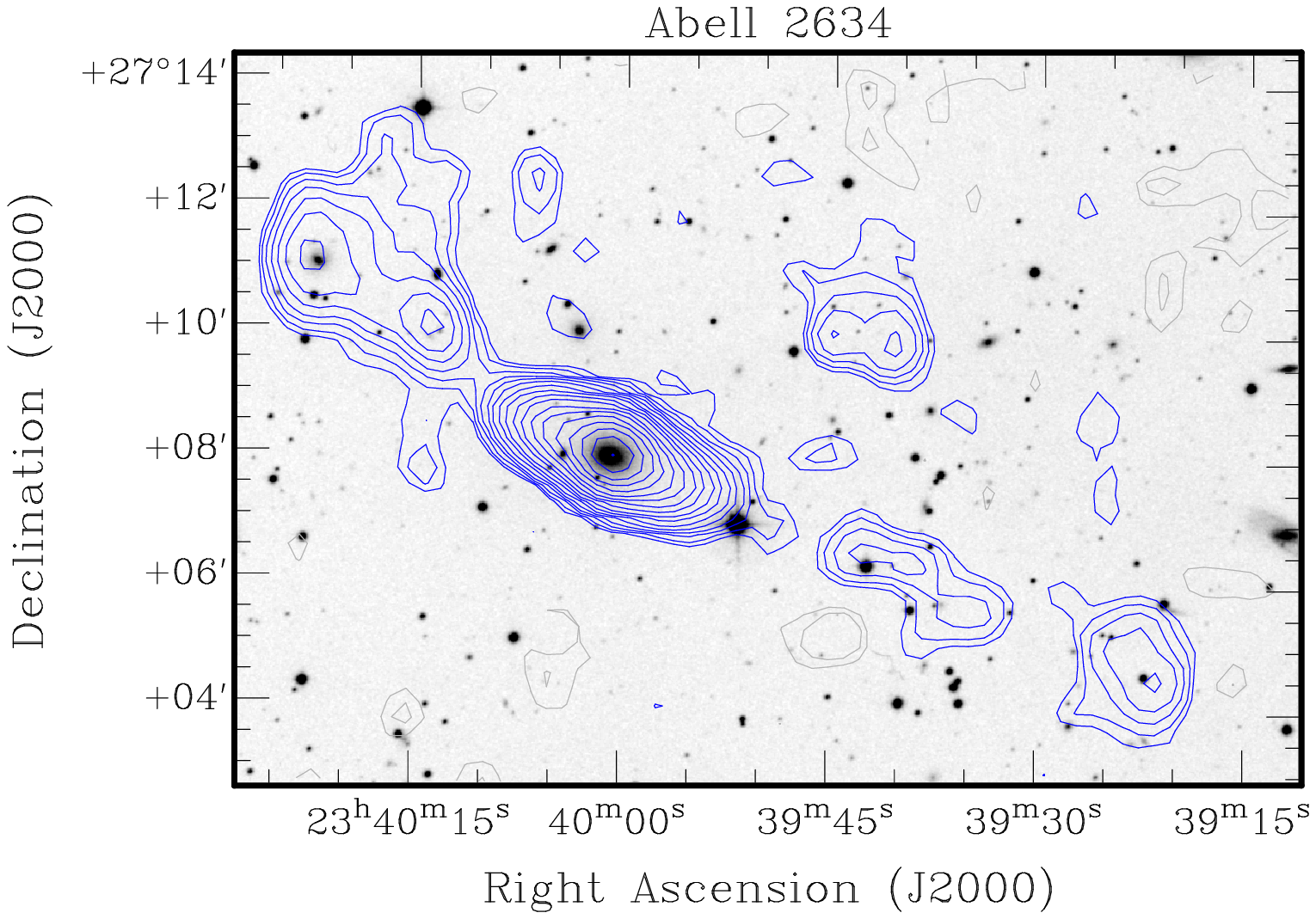}
\caption{NVSS contours overlaid on a red DSS2 image for NGC\,7728 (at J234000.8+270801), the 
2nd-brightest galaxy in A2634 ($z$=0.0317). Coordinates are J2000, and negative
contours are in grey. The radio structure
is suggestive of a single FR\,I/II source of LAS$\sim$16$'$ 
or LLS$\sim$570\,kpc. Since the first, tentative association of such a 
large radio galaxy with NGC\,7728 by \cite{WIEL78} no other large-scale
radio image more sensitive than the NVSS has been published.
Available VLA archive data do not permit to decide whether the cluster member
at J234022.1+271105 is responsible for the apparent NE outer lobe of NGC\,7728.
}
\label{fig1}
\end{center}
\end{figure}

From the major radio surveys currently available (NVSS, SUMSS, FIRST, 
VLSS, WENSS and WISH), we extracted images of 15$'$ on a side around
each BCM. For FIRST we used an image size of 3$'$, given its higher 
angular resolution and lower sensititivy for extended structure.
Only four nearby sources turned out to be larger than 15$'$ and required
larger images (see e.g.\ Fig.~\ref{fig1}).
By visual inspection of $\sim$3000 survey maps, with the BCM marked by a 
cross, we assessed the identification, extent and, where possible, the
morphology of radio sources associated with these BCMs. Radio-optical
overlays were prepared for 270 images in order to check
whether neighboring radio components pertained to the BCM or were more 
likely associated with unrelated optical objects on the red 2nd-epoch 
Digitized Sky Survey (DSS2, archive.stsci.edu/dss).
Upper limits to the flux density were assigned to undetected BCMs. 
For complex extended sources (150 cases), as well as detectable sources 
below the threshold of the corresponding radio source catalogs (115 cases),
flux densities were integrated with the help of the AIPS package 
(www.aips.nrao.edu).

Using the two major catalog browsers (CATS at cats.sao.ru and VizieR at 
vizier.u-strasbg.fr), the following information was extracted for
each BCM radio source (if detected): position, observing frequency and
angular resolution, flux density and its error, deconvolved angular
size, radio position angle (RPA), and reference.  Additional flux densities
were extracted from the catalog collection of the first author (see \cite{AND09}).  

For the sufficiently resolved sources we used the above survey images 
as well as published high$-$resolution images to estimate the best available 
RPA and largest angular size (LAS) of each source.  Occasionally,  these
high$-$resolution images revealed that the apparent association with
the BCM in the low-resolution radio surveys disappeared in favor of a
fainter optical object (e.g.\ MRC~1103$-$244 in A1165, MRC~0001$-$233
in A2719, FIRST~J115518.6+232422 in A1413).

\section{Results}

We compiled the radio data in a table with one record for each 
reference and observing frequency, often with several records for 
the same frequency. From 289 different references we collected a total 
of 5293 records of which 292 were discarded for various reasons 
(e.g.\ too large beam size, overresolution in VLBI observations,
duplication of other original catalog fluxes, etc.). Of the remaining
5001 entries, 2627 are detections and 2374 upper limits.
Of the 1423 BCMs, we found 578 (41\,\%) in 544 different Abell clusters
with detected radio emission. However, 223 of these 578 are detected
at only one frequency, mostly 1.4~GHz (NVSS or FIRST) or 843~MHz
(SUMSS), and are thus usually faint and unresolved, not allowing one to
estimate their RPA or radio morphology.

\begin{figure}[t!]
\vspace*{-3mm}
\begin{center}
\includegraphics[width=0.62\textwidth,,viewport=36 63 450 460,clip]{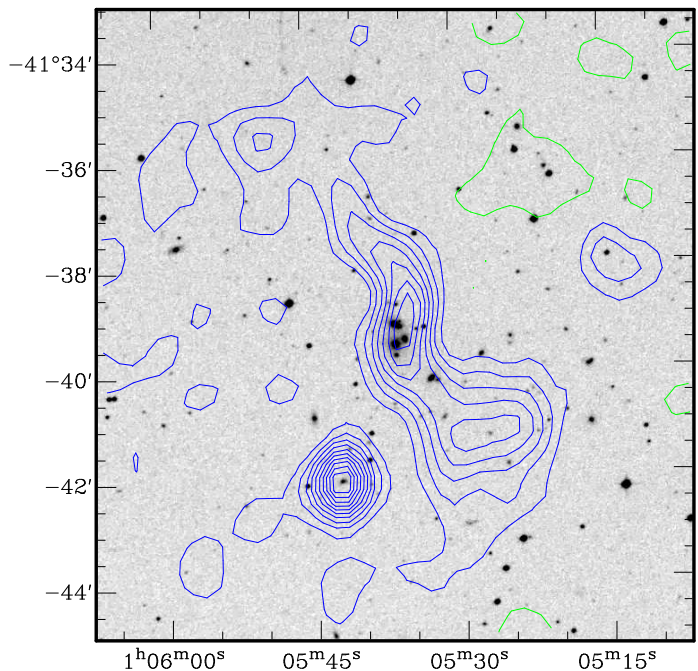}
\caption{SUMSS 843-MHz contours overlaid on a red DSS2 image for S0122
at $z$=0.097, based on 15 members. The central radio galaxy 
has LAS$\sim$9$'$ or LLS$\sim$1\,Mpc.
Of the possible hosts in the central compact group of galaxies only the 
brightest (SE) is a confirmed cluster member, while the 3rd-brightest (due W)
of these is closest to the symmetry center of the radio galaxy at the
given angular resolution of 45$''$.}
\label{fig2}
\end{center}
\end{figure}

Using the BCM redshift (spectroscopic, or else photometric) 
from \cite{COZIOL09} we converted LAS to largest linear size (LLS).
The median LLS of 209 BCMs with available LAS is 160~kpc 
($H_0$ = 75 km\,s$^{-1}$\,Mpc$^{-1}$). We found 29 BCMs
with LLS$>$500\,kpc, including several examples of very large  sources
(LLS$>$700~kpc) in A\,555, A2372, S\,122 (cf.\ Fig.~\ref{fig2}.), S\,527, 
and S\,239, some of them not previously reported. Several new candidate 
wide-angle tailed sources 
were found, e.g.\ in the BCMs of A~555, A~734, A~941, S~250, S~646, and S~793.
Although we did not yet derive radio continuum spectra for all objects,
we noted new examples of very steep-spectrum sources 
($\alpha>$1.4, $S_{\nu}\propto\nu^{-\alpha}$) in the BCMs of
A~122, A~733, A1650, A2110, A2533, A2554, A3497, and S~651, as well
as very flat-spectrum sources ($\alpha\lesssim$0) in the BCMs of
A1644, A2292, A2631, A2660, and A3407.

Looking for stringent upper limits to the 1.4-GHz flux in the nearest
clusters, we found the most radio$-$quiet sources in giant 
ellipticals ($M_B<-$21) with a spectral radio power 
$P_{1.4\,GHz}< 2\times10^{21}\,W\,Hz^{-1}$ in A~779, S~836, and S~900. 
Cross-correlating our BCM sample with the compilation of \cite{MASSA09}, 
we found the BCMs in A3537, A3581, S~549, and S~780, to coincide with blazars.

We could measure the RPA for 183 BCMs, classified into three 
categories of reliability: good, medium and poor, corresponding to errors
of $<5^{\circ}$ (106),  $<10^{\circ}$ (41), and $>10^{\circ}$ (37).

\begin{figure}
\vspace*{-3mm}
\begin{center}
\includegraphics[width=0.6\textwidth,,viewport=5 1 722 508,clip]{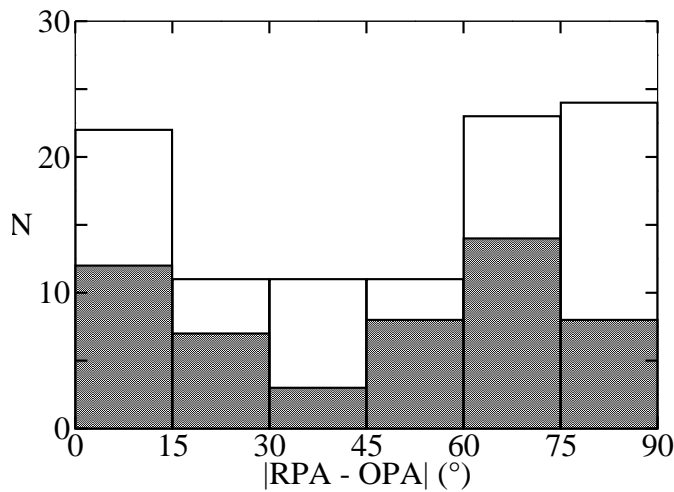}
\caption{Radio$-$optical position angle difference for 102 brightest members of 
Abell clusters. The shaded area represents the best quality data (50 BCMs) with 
$\Delta dPA\lesssim10^{\circ}$, and the white area includes a further
52 BCMs with $\Delta dPA\lesssim15^{\circ}$.}
\label{fig3}
\end{center}
\end{figure}

For the majority of BCMs with not too complex optical structure we
obtained the ellipticity ($\epsilon$) and the major axis position angle
(OPA) of the outermost envelope of the optical galaxy, using the task 
{\tt\small ellipse} in the IRAF package (iraf.noao.edu) on red DSS2
images. These were mostly from \cite{ALAMO07}, but for 181 BCMs (13\%) 
these were determined in the present work.  For 178 BCMs with both RPA and 
OPA we derived the acute difference angle, dPA, between RPA and OPA.
Fig.~\ref{fig3} shows the distribution of dPA's for 102 BCMs with
errors of $\lesssim10^{\circ}$ for both RPA and OPA, and for 50 of these 
for which the data were of even better quality. Fig.~\ref{fig3} shows 
marginal evidence for
a bimodal distribution, with a majority of BCMs showing a perpendicular 
alignment (dPA$\gtrsim 60^{\circ}$) and a smaller fraction ($\sim$20\%) 
a parallel alignment (dPA$\lesssim 15^{\circ}$). However, a 
Kolmogorov$-$Smirnov (KS) test does not reject the null hypotesis of 
a uniform distribution ($P=0.86$ for the good data, and $P=0.46$ for all data).
We also separated our sample by optical morphology of the BCMs
(E- versus D/cD types, taken from \cite{COZIOL09}), by their optical 
ellipticity ($\epsilon>0.2$), 
and the LLS of the radio sources (larger or smaller than 200\,kpc,
as suggested by \cite{PALI79}), and found that none of these parameters
has any influence on the distribution of the radio-optical misalignment 
angle dPA.

\section{Conclusions and Outlook}

Despite the fact that the current study was based on the largest
sample of BCMs in Abell clusters ever compiled \cite{COZIOL09}, the
number of 102 BCMs with available radio-optical misalignment angle  
obtained here, is lower than in the previous sample \cite{AND95},
which was derived by cross-identifying extended radio sources
with brightest cluster galaxies in ANY type of Abell cluster, i.e.\ without
the restrictions used by \cite{COZIOL09}. Given their large overlap,
the shape of the distribution of radio-optical alignment angles in these
two samples is very similar, and suggests a bimodality with a small fraction of
aligned sources and a larger fraction of sources with radio axes
roughly along their optical minor axis. However, KS tests show that 
the distributions of dPA in both the present sample and the larger one of 
\cite{AND95} are indistinguishable from a uniform one.

Thus, at least based on the present study of radio-optical alignments
of brightest cluster members in low-redshift clusters, there is little or 
no evidence for the anisotropic merger scenario proposed in \cite{WEST94}.
Moreover, a selection of Abell clusters with dominant galaxies, based on
Bautz-Morgan or Rood-Sastry type of the clusters alone, seems too restrictive.
In fact, during the visual inspection of DSS images by \cite{COZIOL09} 
it became clear that not all clusters with ``early'' BM types (I and I-II) 
really have a dominant galaxy, while an inspection of DSS images
of 869 Abell clusters of type BM~II \cite{RAMOS06}
revealed that two thirds of them still have a reasonably dominant galaxy.

We also note that the definition of the BM type is not entirely consistent 
in the literature. Even the original definition of the BM~I type
\cite{BM70} varies from ~{\it ``Clusters containing a
supergiant D galaxy''}~ in their abstract, to clusters ~{\it ``containing
a centrally located cD galaxy''}~ in their Table~1a. In a sequel paper, Bautz
\cite{BAU72} insisted that ~{\it ``For
the cluster to be assigned as type~I, the cD galaxy must stand out in
optical appearance from the rest of the cluster to the degree shown by
the brightest members of A2199, A2029, and A2670.''}~
Abell et al.\ \cite{ACO89} who visually classified all clusters on the BM system,
stated that ~{\it ``The BM system was used in its
original form where the magnitude difference between the first and second
brightest galaxies in the cluster is the major classification criterion.''}~
It turns out that 80\% of the present sample of 1169 clusters are in the south
($\delta_{2000}<0^{\circ}$), while only 59\% are expected if BM types
were independent of sky position, given that there are 2167 and 1909 A-clusters
with $\delta_{2000}>0^{\circ}$ and $\delta_{2000}<0^{\circ}$, plus 1174
S-clusters with $\delta_{2000}<0^{\circ}$. This, and the fact that the present
cluster sample has 36\% S-clusters compared to 22\% in the entire Abell
catalog, shows a significant bias towards earlier BM classes in the
southern extension of the Abell catalog as compared to its earlier
northern part.


Clearly, for a significant enlargement of the present sample, a much
larger fraction of Abell clusters needs to be examined for dominant
galaxies, preferably with additional use of photometric magnitudes of the
BCM candidates to assess quantitatively their dominance in the cluster.

Nevertheless, the current compilation provides the basis for a number of future studies,
like e.g.\ the determination of the radio spectral index and shape, and its
relation with the density of the intracluster medium (from data on X$-$ray
emission), the determination of the spectral and total radio luminosity
of BCMs, and a comparison with field elliptical galaxies. It will also
be possible to study relations between optical ellipticity and radio
power or luminosity, as well as between the peculiar velocity of the BCMs within
their clusters (already derived by \cite{COZIOL09}) and the distortion
of their radio morphology, using references from the present compilation.

\acknowledgments
We benefitted from grant 50921-F of Mexican CONACyT, and from the hospitality
at AIfA Bonn, Germany, where part of this research was done.
HA is grateful for additional support from CONACyT grant 81356, as well as
from German DFG through grants RE1462/2 and TRR33.

\end{document}